\documentclass{egpubl}

\usepackage{eg2021}

\Preprint

\usepackage[T1]{fontenc}
\usepackage{dfadobe}  
\usepackage{flushend}

\electronicVersion
\usepackage[pdftex]{graphicx} \pdfcompresslevel=9

\usepackage{xcolor}
\definecolor{lloydColor}{HTML}{FF7744}
\definecolor{jitterColor}{HTML}{4477FF}

\usepackage[caption=false,font=footnotesize,labelfont=sf,textfont=sf]{subfig}
\usepackage{graphicx}
\usepackage{enumitem}
\usepackage{booktabs}
\usepackage{algorithm}
\usepackage{algpseudocode}
\usepackage{hyperref}
\usepackage{xspace}
\usepackage{comment}

\usepackage{graphicx}
\usepackage{soul}
\usepackage{amsmath}
\usepackage{amssymb}
\usepackage{microtype}
\usepackage{wrapfig}
\usepackage{xcolor}
\usepackage{booktabs}
\usepackage{multirow}
\usepackage{xspace}

\def\figurePath{./}

\def\myfigure#1#2{%
    \begin{figure}[htb]%
    \centering\includegraphics*[width = \linewidth]{\figurePath#1}%
    \vspace{-.2cm}%
    \caption{#2}%
    \label{fig:#1}%
    \end{figure}%
}

\newcommand{\mywfigure}[3]{%
\begin{wrapfigure}{r}{#2\columnwidth}%
 \vspace{-.2cm}%
  \begin{center}%
    \includegraphics[width=#2\columnwidth]{\figurePath#1}%
    \vspace{-.2cm}%
    \caption{#3}%
    \label{fig:#1}%
    \vspace{-.2cm}%
  \end{center}%
\end{wrapfigure}%
}

\newcommand{\eg}{e.\,g.,\ }
\newcommand{\ie}{i.\,e.,\ }
\newcommand{\etal}{et~al.\ }

\newcommand{\argmin}[1]{\underset{#1}{\operatorname{arg\,min}\ }}

\newcommand{\refSec}[1]{Sec.~\ref{sec:#1}}
\newcommand{\refFig}[1]{Fig.~\ref{fig:#1}}
\newcommand{\refEq}[1]{Eq.~\ref{eq:#1}}

\newcommand{\refAlg}[1]{Alg.~\ref{alg:#1}}

\newcommand{\change}[1]{#1}

\soulregister\ref7
\soulregister\cite7
\soulregister\refFig7
\soulregister\refAlg7
\soulregister\cite7
\soulregister\ref7
\soulregister\pageref7
\soulregister\shortcite7
\soulregister\eg0
\soulregister\ie0
\soulregister\etal0

\DeclareGraphicsExtensions{.png,.jpg,.pdf,.ai,.psd}
\newcommand{\mysection}[2]{\section{#1}\label{sec:#2}}
\newcommand{\mysubsection}[2]{\subsection{#1}\label{sec:#2}}
\newcommand{\mysubsubsection}[2]{\subsubsection{#1}\label{sec:#2}}

\newcommand{\mymath}[2]{\newcommand{#1}{\TextOrMath{$#2$\xspace}{#2}}}

\newcommand{\name}{Blue Noise Plot\xspace}
\newcommand{\names}{Blue Noise Plots\xspace}

\mymath{\inputValue}{x}
\mymath{\outputValue}{y}
\mymath{\stackedPoint}{\mathbf p}
\mymath{\voronoiPoint}{\mathbf q}
\mymath{\inputValues}{X}
\mymath{\outputValues}{Y}
\mymath{\stackedPoints}{P}
\mymath{\numberOfPoints}{n}
\mymath{\distribution}{f}
\mymath{\numberOfIterations}{m}
\mymath{\cell}{V}
\mymath{\cellCenter}{C}
\mymath{\metric}{\kappa}
\mymath{\density}{d}
\mymath{\height}{h}
\mymath{\period}{L}

\newcommand{\dataset}[1]{\texttt{#1}}

\newcommand{\method}[1]{\textsc{#1}}
\newcommand{\methodOur}{\method{Our}\xspace}
\newcommand{\methodJitter}{\method{Jitter}\xspace}

\newcommand{\mypara}[1]{\noindent\textbf{#1}\ }

\title{\names}

 \author[C. van Onzenoodt, G. Singh, T. Ropinksi, T. Ritschel]
 {
 $^1$Christian van Onzenoodt\qquad 	
 $^2$Gurprit Singh\qquad
 $^1$Timo Ropinski\qquad
 $^3$Tobias Ritschel
 \\
 $^1$Ulm University\qquad		
 $^2$Max-Planck Institute for Informatics, Saarbr\"ucken\qquad	$^3$University College London
 }

\teaser{
  \centering
  \includegraphics[width=\textwidth]{Teaser.ai}%
  \caption{\names \textbf{(left)} prevent clutter and provide visually more appealing results than frequently used jitter plots \textbf{(right)}.
  Importantly, \names are \emph{unbiased}, in the sense that no data point is ever changed and  strictly all points of the sample are presented.}%
  \label{fig:teaser}
}%

\begin{document}
\maketitle

\begin{abstract}
We propose \names, two-dimensional \change{dot plots that depict data points of univariate data sets.}
While often one-dimensional strip plots are used \change{to depict such data}, one of their main problems is visual clutter which results from overlap.
To reduce this overlap, jitter plots were introduced, whereby an additional, non-encoding plot dimension is introduced, along which the data point representing dots are randomly perturbed. Unfortunately, this randomness can suggest non-existent clusters, and often leads to visually unappealing plots, in which overlap might still occur.
To overcome these shortcomings, we introduce \names where random jitter along the non-encoding plot dimension is replaced by optimizing all dots to keep a minimum distance in 2D \ie Blue Noise.
We evaluate the effectiveness as well as the aesthetics of \names through both, a quantitative and a qualitative user study.
The Python implementation of \names is available \href{http://github.com/onc/BlueNoisePlots}{\texttt{here}}.
\end{abstract}

\mysection{Introduction}{Introduction}
Consider depicting a univariate data set, \eg observed ages in a cohort, on paper.
While we could simply report first order statistics, such as for instance the mean, this would be an oversimplification in many important cases~\cite{taleb2007black}.
Instead, we would like to show all data points, and thus ask for the optimal way to represent these with dots in a two-dimensional plot.

A \emph{strip plot} simply plots data points on a single horizontal axis, while an observer is free to apply domain knowledge to judge what is the density, what might be modes or what might be outliers.
Recent efforts, to communicate data to a wider public and not just to experts have resulted in such plots to be used increasingly in print media, television and on the web.
Strip plots are most effective when the number of data points is still low enough to be displayed, but high enough to represent the information.
The first row of \refFig{Concept} shows examples of strip plots on a univariate toy data set. Unfortunately, the main disadvantage of using strip plots to depict such data sets, is \emph{clutter} which often leads to overdraw. As such, if two dots $\inputValue_i$ and $\inputValue_j$ are closer than what the printer, display or the human visual system can reliably discern, the advantage is lost, since not all data points are effectively communicated. For instance, the data set $\{1,2,4\}$ results in the same visual representation, as the data set $\{1,2,2,4,4\}$, because the individual dots representing the values $2$ and $4$ would not be discernible when plotting the second data set, due to overdraw.

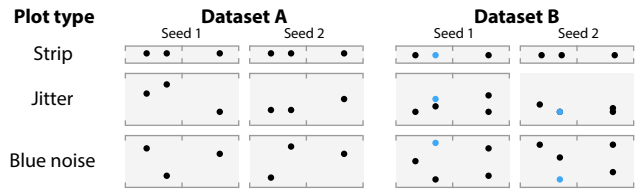
\myfigure{Concept}{Three different plotting approaches \textbf{(rows)} for two different univariate datasets at two different random seeds \textbf{(columns)}.
Strip plots always fail to convey the datasets, as they look the same.
Jitter plots, may sometimes fail: dataset B looks like dataset A for seed 2 while they are different.
Our \name never fails.}

As a remedy, \emph{jitter plots} have been proposed~\cite{chambers2018graphical,tukey1977exploratory}, which introduce an additional, non-encoding dimension, along which dots are randomly perturbed. For our example, of univariate data points plotted in two dimensions, this means to simply move the respective dots vertically by a random amount. The second row of \refFig{Concept} shows examples of such jitter plots on the before data sets. While jitter plots usually reduce the amount of clutter and are easy to implement, they have three main drawbacks.
First, the introduced randomness leads to gaps and clusters, which might be falsely perceived as features present in the data set.
Second, no minimum distance between dots is enforced, which in the worst case might lead to overlap, something we have observed in many real world examples.
Lastly, jitter plots do often look visually unappealing.

By introducing \names, we provide a solution to the three main drawbacks of jitter plots. A conceptual comparison is made in the last row of \refFig{Concept}.
Computer graphics has explored \emph{blue noise}~\cite{ulichney1988dithering}, that is, dot patterns, that are still random, but without dots getting to close to each other. By proposing a modified Loyd relaxation algorithm~\cite{lloyd1982least}, we can extend jitter plots to become \names. Importantly, \names are \emph{unbiased}, in the sense that we make no assumption on how data points are represented, no data point is ever changed and strictly all data points of a data set are presented, setting it apart from other methods~\cite{bertini2004chance,bachthaler2008continuous,keim2010generalized,mayorga2013splatterplots,micallef2017towards}, that either re-sample or change the way data points are presented.
  
To investigate the impact of \names on task performance and visual appeal, we have conducted a crowd sourced user study, whereby the obtained results indicate, that \names are beneficial over conventional jitter plots. Furthermore, we have performed a qualitative study assessing the visual appeal of \names, and discuss objective quality measures.

\mysection{Previous Work}{PreviousWork}
Our work addresses the visualization problem of plotting univariate data sets using the computer graphics methodology of blue noise, both of which we will review now. 

\mypara{Visualizing univariate data sets.} To visualize univariate data sets, several different plot types exist. We distinguish between direct and aggregated visualizations. While direct methods allow for the depiction of the individual data points forming the data set, aggregated methods only communicate the data set per se, often in an approximate manner. Several well known plots fall into the later category, such as for instance histograms~\cite{pearson1894contributions}, \emph{box plots}~\cite{spear1969practical}, and \emph{violin plots}~\cite{hintze1998violin}. Nevertheless, this category is not in the focus of our work, and we instead aim for direct techniques, which explicitly convey the existing data points. One of these techniques, that is frequently used are strip plots~\cite{cleveland1985elements}, where data points are represented as symbols, usually dots or circles. As stated above, these symbols are simply plotted all along the same dimension, independent of the occurring densities. While strip plots are an intuitive way to convey univariate data sets in a direct manner, they come with severe limitations, as they suffer from overdraw in dense regions. As a consequence, they might introduce distortion, as the maximum density to be communicated is limited by this overdraw. There are two main approaches to deal with the problem of overdraw. The size and or alpha blending value of the representing symbols can be altered, or data points can be transformed such that overdraw is reduced. While naturally, these approaches can be combined, we solely focus on transformation approaches in this paper. 

Among the techniques applying transformations, the \emph{stacked dot plot} is the most basic, as it simply stacks the shown dots, which effectively resembles a bar chart for non-continuous data~\cite{cleveland1993visualizing,wilkinson1999dot}. For continuous data sets, more elaborate packing schemes need to be employed, in order to obtain an acceptable layout. So-called \emph{beeswarm plots} exploit such packing which enables them to obtain a stacked and dense representation without the need for binning, but their simple construction leads to false features and clustering.

In contrast to the stacked representations described above, jitter plots randomly distribute all points in a given range along an additional, non-encoding dimension~\cite{tukey1977exploratory}. Thus, it also becomes possible to reduce the amount of overdraw, to a degree dependent on the plot size and the occurring densities. Jittering originally goes back to Chamber et al.~\cite{chambers2018graphical}, while Tukey and Tukey additionally exploit constraints~\cite{tukey1990strips}.

Recently, several visualization approaches have been proposed, which deal with the shortcomings of dot plots in general. Bachthaler and Weiskopf have introduced continuous scatter plots, which sacrifice the discrete nature of scatter plots, in order obtain a dense visualization~\cite{bachthaler2008continuous}. Mayorga and Gleicher go further by automatically grouping some dots, while keeping others~\cite{mayorga2013splatterplots}, which is combined with interactive exploration. Along a similar line, Bertini and Santucci introduce non-uniform samplings in order to communicate density in 2D scatter plots~\cite{bertini2004chance}. Keim et al. introduce generalized scatter plots, where the dot locations are modified in order to reduce overlap~\cite{keim2010generalized}, which in contrast to our approach, this adds bias to the actual data values. Micallef present an optimization approach for the perceptual optimization of scatter plots~\cite{micallef2017towards}. By exploiting task-dependent cost functions, they are able to obtain satisfying scatter plot designs, which compete with those crafted by humans.
\change{Yuan~et~al.\cite{yuan2020evaluation} and Rapp et al. \cite{rapp2020void} also address the problem of overdraw in scatter plots where they select a subset of data points from a large data set such that the resulting patterns follows the density, yet has blue noise.
Our work does not select subsets, but shows all data points, and introduce an additional, non-encoding dimension so that dots can become blue noise in the first place.}

Some forms of box plot also communicate individual data points, e.g., when they are outliers, and thus can be considered a combination of aggregated and direct visualization. Many of the contributions made in this paper, can also be applied in this context.

%A notable application of blue noise to a visualization problem is presented by Hu\etal\cite{hu2019data}.
%They assume access to multiple distributions they can sample, and do so, such that the samples are maintaining a blue noise relation.
%Please, note that even their special case of a single class is a fundamentally different problem: Their method is free to sample a given distribution, while we have to make do with samples where no points are added or removed, leading to a novel optimization problem of partially-constrained blue noise.

\begin{figure*}
\begin{minipage}{0.31\textwidth}
\begin{algorithm}[H]
    \color{lloydColor}
    \caption{Lloyd relaxation}
    \label{alg:Lloyd}
    %\centering
    \begin{algorithmic}[1]
    \State $\stackedPoints \leftarrow \mathtt{uniform}()$
    \Repeat
        \State $V = \mathtt{voronoi}(\stackedPoints)$ 
        \For{$\stackedPoint_i\in\stackedPoints$}
        \State $\stackedPoint_i \leftarrow 0$
        \For{$\voronoiPoint_j\in V_i$}
            \State $\stackedPoint_i \leftarrow \stackedPoint_i + \voronoiPoint_j / |V_i|$
        \EndFor
        \EndFor
        \State\hphantom{$X$}
    \Until{converged}\\
    $\mathtt{plot}(\stackedPoints)$
    \end{algorithmic}
\end{algorithm}
\end{minipage}
\hfill
\begin{minipage}{0.02\textwidth}%
\vspace{0.7cm}
+
\end{minipage}
\hfill
\begin{minipage}{0.31\textwidth}
\begin{algorithm}[H]
    \centering
    \color{jitterColor}
    \caption{Jitter Plots}
    \label{alg:JitterPlots}
    \begin{algorithmic}[1]
    \State $\stackedPoints \leftarrow \mathtt{stack}(\inputValues, \mathtt{uniform}())$
    \\
    \\
    \\
    \\
    \\
    \\
    \\
    \\
    \\
    \\
    \\
    $\mathtt{plot}(\stackedPoints)$
    \end{algorithmic}
    \end{algorithm}
\end{minipage}
\hfill
\begin{minipage}{0.02\textwidth}%
\vspace{0.7cm}
=
\end{minipage}
\hfill
\begin{minipage}{0.31\textwidth}
\begin{algorithm}[H]
    \caption{Blue Noise Plots}
    \label{alg:BlueNoisePlots}
    \begin{algorithmic}[1]
    \color{jitterColor}
    \State $\stackedPoints \leftarrow \mathtt{stack}(\inputValues, \mathtt{uniform}())$
    \color{lloydColor}
    \Repeat
        \State $V = \mathtt{voronoi}(\stackedPoints)$ 
        \For{$\stackedPoint_i\in\stackedPoints$}
        \State $\stackedPoint_i \leftarrow 0$
        \For{$\voronoiPoint_j\in V_i$}
            \State $\stackedPoint_i \leftarrow \stackedPoint_i + \voronoiPoint_j / |V_i|$
        \EndFor
        \EndFor
    \color{black}
    \State $\stackedPoints \leftarrow \mathtt{stack}(\inputValues, \mathtt{unstack}(\stackedPoints, \inputValues))$
    \color{lloydColor}
    \Until{converged}
    \color{jitterColor}
    \\
    $\mathtt{plot}(\stackedPoints)$
    \end{algorithmic}
    \end{algorithm}
\end{minipage}
\caption{Our approach \textbf{(right)} is a combination of Lloyd relaxation \textbf{(left)} and jitter plots \textbf{(middle)} with a data constraint extension.
The function $\mathtt{stack}(A,B)$ stacks vector $A$ on top of vector $B$.
The function $\mathtt{unstack}(A,B)$ returns the vector $A$ with the dimensions from $B$ removed. 
}
\end{figure*}

\mypara{Blue noise.}
Dot patterns are often described in terms of their expected power spectrum profile. Patterns exhibiting mostly high-frequency content in their power spectrum are characterized as \emph{blue noise}. The resulting spatial distribution of dots respect some minimum distance giving perceptually pleasing patterns~\cite{yellott1983spectral}. Consequently, blue noise has been widely adopted in  many computer graphics applications including halftoning~\cite{ulichney1988dithering}, stippling~\cite{secord02weighted}, artistic packing~\cite{reinert13interactive}, anti-aliasing~\cite{dippe85anti} and variance reduction for Monte Carlo rendering~\cite{singh19analysis}.
Typically, such methods are free to place dots in whatever arrangement, only their correlation and in some cases an importance function is relevant.
Also the dot count is typically not fixed strictly.
An exception is the work by Reinert et al.~\cite{reinert13interactive}, who initialize Lloyd relaxation with coordinates that are correlated with a feature vector of a specific set of elements.
In comparison to our approach, this is biased, \ie dots are free to change position or order, defying the purpose of a jitter plot altogether, which is to link individual dots to definitive data points.
Our work shares similarity with Reinert~\etal\cite{reinert2015projective} that treat different dimensions in the dot optimization differently, but without complying to data points, only optimizing for different spectra.
In this work, we show that dots distributed while obeying to blue noise provide better data visualizations. Therefore, we optimize the dot layout using Lloyd relaxation~\cite{lloyd1982least} to obtain a blue noise distribution.
However, unlike traditional approaches, our optimization works by keeping the encoding dimension fixed, while optimizing the dot positions along the other, non-encoding dimension.
Our optimization runs in higher dimension than the data and uses a novel distance metric to emphasize the non-encoding dimension to guide the optimization.
We already have mentioned Hu~\etal~\cite{hu2019data} as a rare example of a visualization paper that relates to blue noise.
Their task might be easier than ours, as they assume sampling from multiple importance functions to produce multi-class blue noise, without adhering to ordinality or coordinates.

\mysection{Blue Noise Plots}{OurApproach}
We will here give a formal definition of our approach, starting from in- and output (\refSec{InputOutput}), to a variational formulation with constraints (\refSec{CostFunction}) and our implementation to minimize it (\refSec{Optimization}).
The section concludes with several extensions, such as centrality, resembling bee swarm plots, automatically choosing the plot height (\refSec{AutomaticHeight}), as well as introducing a multi-class version (\refSec{Multiclass}).

\mysubsection{Input and Output}{InputOutput}
\change{
Input to our method is a set $\inputValues=\{\inputValue_i\in\mathbb R\}$ of univariate data points, where values are assumed to be associated with the horizontal axis.
Output is a set $\outputValues=\{\outputValue_i\in\mathbb R\}$ of scalar jitter values associated with the vertical axis.
We write
$
\stackedPoints = 
\{
\stackedPoint_i=(\inputValue_i,\outputValue_i)
\in\mathbb R^2
\}
$ 
to denote the combination of data values along the horizontal (encoding) dimension, and jitter value along the vertical (non-encoding) dimension into a set of two-dimensional dots.
}

\mysubsection{Cost Function}{CostFunction}
Optimization is performed for \change{the set of vertical jitter values  $\outputValues$ of an} output \name, given the univariate input data set \inputValues.

We minimize the cost
\begin{equation}
\label{eq:Main}
\argmin\outputValues
\sum_{\outputValue_i\in\outputValues}
\mathbb E_{\voronoiPoint\sim V(\stackedPoint_i,\stackedPoints)}
\metric
(\stackedPoint_i,\voronoiPoint),
\end{equation}
the sum of the expected value $\mathbb E$ of the $\metric$-distance from the $i$-th \change{output dot $\stackedPoint_i$ to all sites $\voronoiPoint$ in its Voronoi cell $\cell(\stackedPoint_i,\stackedPoints)$ in respect to all other dots $\stackedPoints$}.

\mypara{Uniform metric.}
For classic \names, we use 
\begin{equation}
\metric(\stackedPoint_1,\stackedPoint_2)=
||
(\stackedPoint_1-\stackedPoint_2)^\mathsf T
\begin{pmatrix}
2 & 0 \\
0 & 1 \\
\end{pmatrix}
||_1.
\end{equation}%
Here, the constant diagonal matrix, emphasises the non-encoding dimension along the vertical direction.
We will in \refSec{AutomaticHeight} introduce further metrics to realize other variants.

\mysubsection{Optimization}{Optimization}
While \change{two-dimensional} Lloyd relaxation~\cite{lloyd1982least} would minimize the distance cost, it unfortunately does not adhere to the hard constraints.
In Lloyd relaxation, after a random initialization (\refAlg{Lloyd}, Line 1) every dot is iteratively replaced by its Voronoi cell center (\refAlg{Lloyd}, Line 2 to 10), followed by a re-computation of the Voronoi cells (\refAlg{Lloyd}, Line 3) in a expectation-maximization procedure~\cite{dempster1977maximum}.
Running it directly, will loose the information present in the data sets, such as trivially done by jitter plots (\refAlg{JitterPlots}).

Our main contribution is a solver that extends Lloyd relaxation to produce dot patterns with even distribution that adhere to the data, as explained in (\refAlg{BlueNoisePlots}). Including the hard constraint is intuitive: use both dimensions for the relaxation (\refAlg{BlueNoisePlots}, Line 1) but never update the encoding dimension (\refAlg{BlueNoisePlots}, Line 10).
\change{During optimization}, dots move \change{vertically, with a single degree of freedom per dot, but } their cost computation, including the Voronoi construction, involves 2D.

Note, how this is different from optimizing only an 2D dot pattern and not involving the 1D information. The information is not updated, but it is crucial to include it in the distance computation, such as to satisfy the Lloyd objective in what is perceived: 2D space.

\myfigure{Voronoi}{\protect\change{Lloyd relaxation involves Voronoi cells $\cell$ (blue, grey and orange areas) computed on the output dot pattern $\stackedPoints$.
These $\cell$ are sampled using sites $\voronoiPoint$ shown here as rectangles.
The optimization relates every dot $\stackedPoint_i$ to all the $\voronoiPoint_j$ in its cell $V(\stackedPoint_i, \stackedPoints)$.}}

\change{In practice, different options to construct Voronoi regions exist.
We follow the approach from Balzer et. al~\cite{balzer2009capacity} but without the capacity constraint. We use $8,192$ random 2D points $\voronoiPoint$ to discretize the domain. Note that faster GPU methods exist, that make use of regular grids~\cite{hoff1999fast,li2010fastccvt}.}

\mysubsection{Automatic Height}{AutomaticHeight}
The output dots \stackedPoints are from a domain that typically is wider than high when using the horizontal axis as the encoding axis, and the vertical axis as the non-encoding dimension.
This is, as most datasets have only a couple of different data points per interval in the data dimension, compared to the total number of data points.
Unfortunately, it might not be obvious how to \change{choose an appropriate height}, both for jitter plots, as well as for \names for a given data set.
\change{
The desired distance between dots is part of the problem definition.
Making recommendations how to choose the distance of two dots such that they become visually discernible is clearly an often-encountered visualization challenge, but out-of-scope for this work. 
We will assume it to be known and use a distance of two-times the dot radius in all results we show.
}
We will now show, how to choose \change{the plot height} automatically and optimally.

\mywfigure{Automatic}{0.5}{Auto-height (see text).}
We assume access to the \emph{density} function $\density(\inputValue)\in\mathbb R\rightarrow\mathbb R^+$, defined on the domain of the data distribution (black in \refFig{Automatic}).
Typically, we are only given a sample of the distribution not the true density.
Hence the density function needs to be estimated from the sample, for which we use Kernel Density Estimation in practice.
Please note, that we do not in general rely on density estimation, unless the plot height is chosen automatically or we optimize for centrality (\refSec{Centrality}).

The optimal height depends on the maximal density $\density_\mathrm{max}$, the total number of data points $\numberOfPoints$, and the desired distance $r$ between dots.
The desired distance is chosen by the user (pink in \refFig{Automatic}).
It depends on the output medium, whereby a typical choice is to make it twice the size of a dot, so they become discernable.
We note that at the data coordinate with $\density_\mathrm{max}$, we need to ``stack'' $r\cdot\density_\mathrm{max}\cdot\numberOfPoints$ dots (orange in \refFig{Automatic}).
Ignoring optimal packings with efficiency around 0.9, and assuming a conservative efficiency of 1 instead, stacking such dots with radius $r$ needs to be $r^2\cdot\density_\mathrm{max}\cdot\numberOfPoints=\height$ (green, \refFig{Automatic}).

\mysubsubsection{Centrality}{Centrality}
The point density in a \name might vary.
Alternatively, we can restrict the points to not use all the space available due to the non-encoding dimension.
To this end, for a fixed width (manual or automatic), we limit dots to move less along the non-encoding dimension, resembling the appearance of beeswarm plots.
We refer to such a plot as having \emph{centrality}.

Choosing a varying height is based on the generalization of the aforementioned automatic height (\refSec{AutomaticHeight}).
Instead of choosing a single height value \height, we choose height as a function $\height(\inputValue)$ of the data coordinate \inputValue itself. 

This idea is conveniently realized by changing the metric itself to be non-uniform. How distant two dots are, is depending on the density at the \change{coordinate $\inputValue_{12}=(\inputValue_1-\inputValue_2)/2$} between them:
\begin{equation}
\metric(\stackedPoint_1,\stackedPoint_2)=
||
(\stackedPoint_1-\stackedPoint_2)^\mathsf T
\begin{pmatrix}
d(\inputValue_{12}) & 0 \\
0 & 1 \\
\end{pmatrix}
||_1
\end{equation}%
Note, that $\stackedPoint_1$ and $\stackedPoint_2$ are \change{typically} close, so even while the function is not a metric for all pairs, it locally is as the density function is smooth. \change{In particular it can be chosen arbitrarily smooth by using a smooth kernel in the density estimation, that, in the limit, corresponds to the constant height.}

\mysubsection{Multi-Class}{Multiclass}
For data sets comprised of different classes, \names can be extended to multi-class blue noise~\cite{wei2010multi}. Here, all data points in one class maximize \refEq{Main}, as well as all possible unions of all data points in all classes do. The solver implements this by alternating between the individual classes and their unions.

\mysection{Results}{Results}
We present both qualitative and quantitative results of our work.

\mypara{Single-class.}
We show results of our, as well as existing approaches on typical data sets in~\refFig{JitterVsBN}. We see, that our approach does minimize the amount of visual clutter in the form of overlap for all examples. While this is  particularly the case in denser regions, even in sparser regions (for example in~\refFig{JitterVsBN}, c) on the right), dots are more evenly distributed over the available domain, supporting the perception of individual dots.

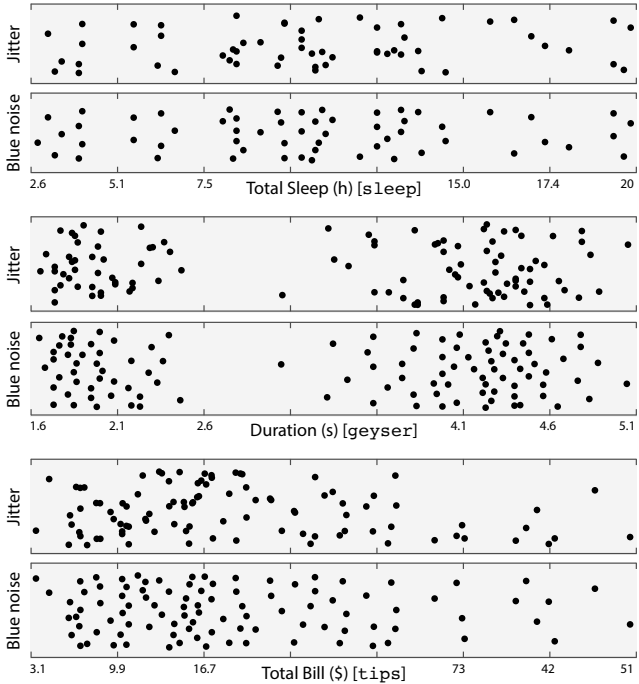
\myfigure{JitterVsBN}{Comparison of three different data sets, each of them visualized using a traditional jitter plot and our \name.}

\mypara{Automatic height.}
As described in~\refSec{AutomaticHeight}, we dynamically adapt the height of the plot, depending on the number of dots and their distribution. 
\change{While \refFig{numPointsConstHeight} shows an example of differently sized subsets of the \protect\dataset{geyser} data set using a constant height for the plot, \refFig{numPoints}, shows the dynamic adaption.}
We do so to compromise between a compact plot and room for the dots to relax and therefore reduce overlap.
If the number of data points gets large, retaining a fixed distance is only possible at the expense of a high plot.

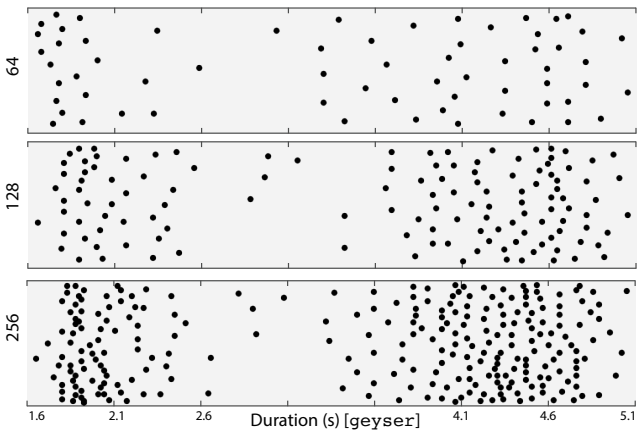
\myfigure{numPointsConstHeight}{Comparison of plots with different numbers of dots. All plots are drawn using the same height, but with different numbers of dots. These plots show a random subset of the \dataset{geyser} data set, visualized using our \name. Here a) shows 64 dots, b) shows 128 dots and finally c) shows 256 dots.}

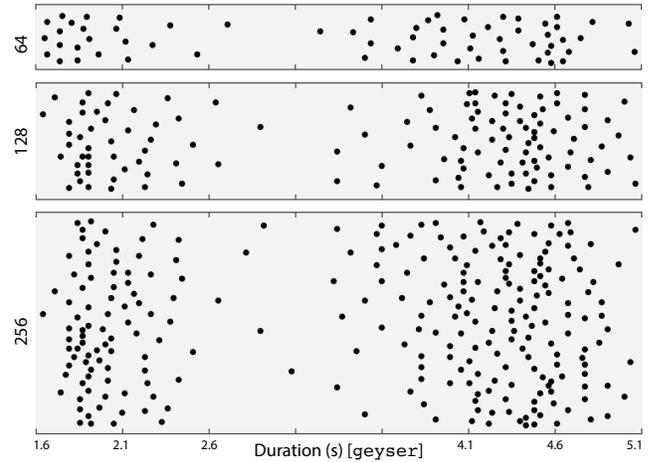
\myfigure{numPoints}{Comparison of adaptive plots with different numbers of dots. These plots show a random subset of the \dataset{geyser} data set, visualized using our \name. Here a) shows 64 dots, b) shows 128 dots and finally c) shows 256 dots.}

\mypara{Centrality.}
\refFig{Centrality} shows results where the height is chosen automatically, but varying with the data dimension. Depending on the reliability of the underlying density estimation, this can be an effective additional cue. At any rate, adding blue noise improves upon jitter in readability and aesthetics.

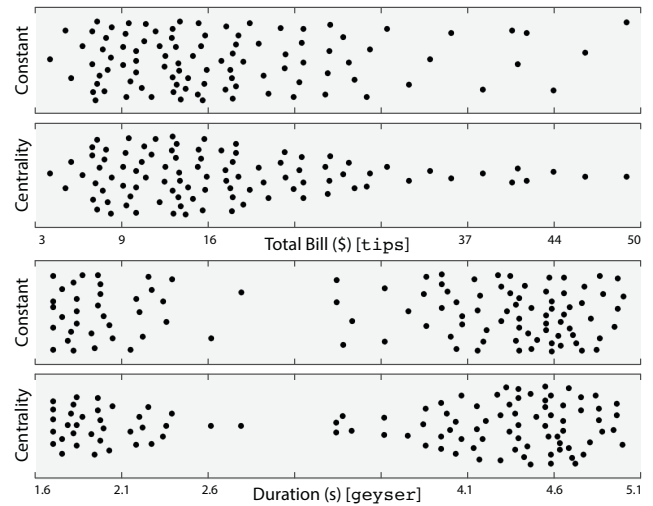
\myfigure{Centrality}{Optimal constant plot height \textbf{(first and third)} and a varying height \textbf{(second and fourth)}, both for \names.}

\myfigure{MulticlassDodge}{Multi-class \name for the \dataset{tips} data set with two classes: \dataset{dinner} and \dataset{lunch} encoded into color.
The first two rows show the individual, the third the combined plot. Please note, how this is three visualizations of \emph{one} set, fulfilling all intra- and inter-class, as well as the data constraints simultaneously.}

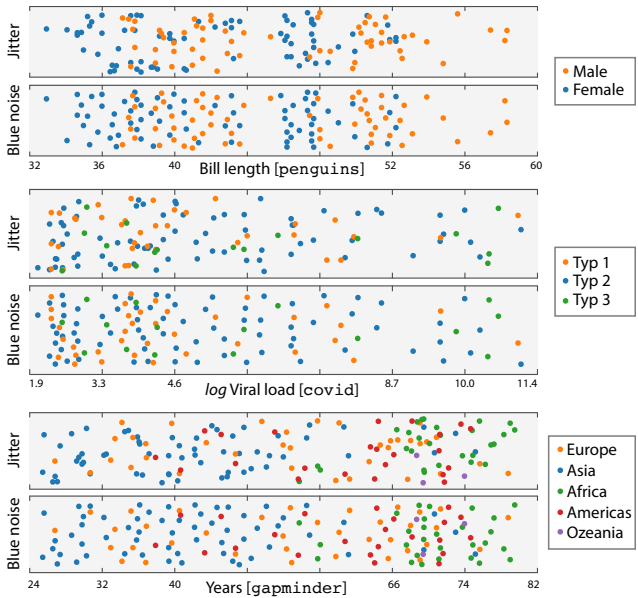
\myfigure{Multiclass}{Different examples of multi-class data sets, visualized using jitter plots as well as \names.
} 

\mypara{Multi-class.}
Finally, we show an extension to multiple classes of data points. Here, every input point additionally has a class label. We use our method to produce a plot that is blue noise for all classes jointly, as well as for every class on its own. \refFig{MulticlassDodge} shows an example of this where the first two rows show the blue noise distributed dots of the individual classes. The third row, shows the final plot, where the first two rows are combined.
\refFig{Multiclass} shows more examples of our \names, encoding multiple classes. Here, we can also see that our approach nicely distributes all the dots, as well as the dots for the individual classes over the entire domain.
Further, \refFig{Multiclass} and \refFig{Quantized} show examples of quantized data.
While the blue noise pattern is less prominent in this case, this shows another strength of our approach. 
In contrast to jitter plots, where the overlap between dots is amplified by this type of data distribution, our approach spreads out the dots vertically.

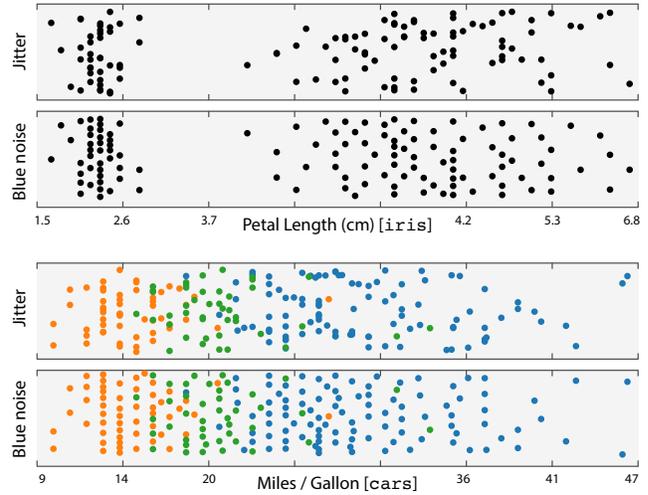
\myfigure{Quantized}{Quantized data sets, such as shown here, where not many $x$ values are shared by a data point, are difficult to optimize for, but worth addressing: besides being \protect\change{less visually appealing in many cases, according to our study}, jitter plots, due to clutter, are \protect\change{more difficult to read}.}

\mypara{Icons.}
Inspired by approaches which represent data points with more complex primitives instead of dots~\cite{hiller2003beyond,reinert13interactive}, we have used our method to place icons as seen in \refFig{Icons}. Here, every data point has a unique icon, making relations visible without bias or clutter.

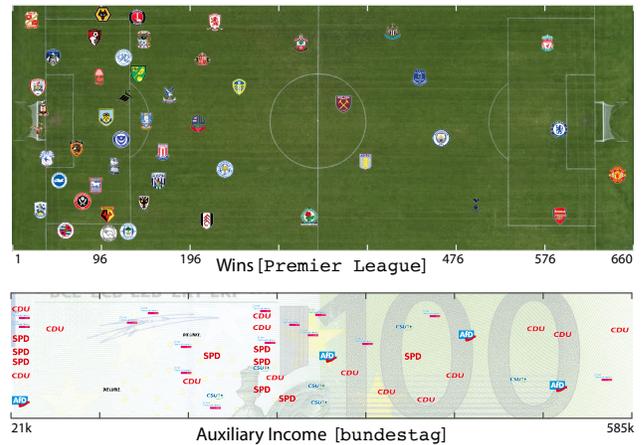
\myfigure{Icons}{\names can further be used to position extended primitives, instead of dots, here, little icons depicting soccer clubs or political parties.}

\mypara{Parameter choice.}
A typical result of a data set containing 256 data points, as shown in this paper, requires 40 iterations of Lloyd relaxation, with 8,192 Voronoi samples, resulting in a total time of six seconds for a naive, non-parallel implementation.

\mypara{Analysis.}
We analyze of our results both from the graphics perspective using the spectral quality of \name as dot patterns, as well as with overlap measures used to analyze plots.

\change{To perform spectral analysis, we compute the expected power spectrum of dots obtained from the} \dataset{geyser} \change{dataset. We generate $100$ different realizations of the dot patterns, compute their power spectrum and average these power spectra to get the expected spectrum. \refFig{Analysis} shows these expected power spectra for} \methodOur (right) and \methodJitter (left).
We compare this against vanilla Lloyd relaxation (middle), which is not a plotting method, as it does not produce an unbiased result, but can serve as some upper bound of what we could achieve when using it as a backbone.
%Results are seen in the bottom of \refFig{Analysis}. 
For \methodJitter, the spectrum is flat like \emph{white noise}.
\methodOur approach gives a dark region in the middle of the spectrum confirming the \emph{blue noise} behavior. \change{The bright line in the middle is due to the non-uniform density of dots along the horizontal axis, where they obey to the data values.}
If we run Lloyd relaxation without constraining the data along the horizontal axis, the dark region in the middle gets larger (middle) \change{and we get uniform density points. That is why, there is no bright line in the middle spectrum.}
\change{The anisotropic structure of the dark region is evident due to the non-square domain of the plot. A domain of length $\period$ has a valid spectrum at only $1/\period$-th frequencies~\cite{singh19fourier}. In~\refFig{Analysis}, since the plot along $x$-axis $\in [0,1)$ and $y$-axis $\in [0,0.2)$, the spectrum is valid only at integer frequencies along the $x$-axis and every ($1/0.2 =$) $5$-th frequency along $y$-axis.
Lastly, the central dark line in the Lloyd relaxation spectrum (middle) implies denser \emph{stratification} of the $x$-axis  w.r.t. the $y$-axis.
}

%The anisotropic structure of this dark region is evident from the non-square domain.
%The radius of the dark region in the spectrum is longer along $x$-axis implying denser \emph{stratification} w.r.t.\ the $y$-axis.

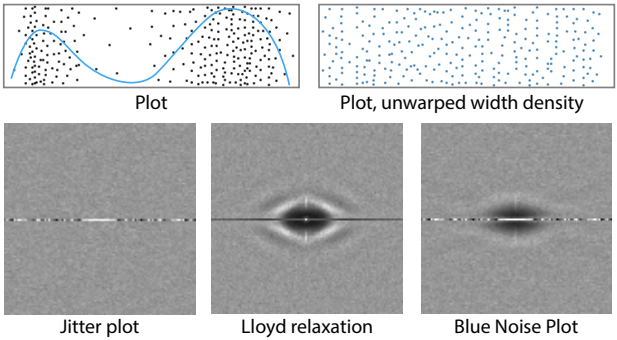
\myfigure{Analysis}{The top part shows the plot, its density function as a blue line and the unwarped plot points.
When performing this on jitter, blue noise based on Voronoi and \name, we find the three spectra seen. A well-distributed dot set should exhibit a low energy (black) in the low-frequency regions (center).
While being inferior to Lloyd relaxation, we fair substantially better than jitter.}

We also analyzed our plots using a point overlap metric presented for scatter plots~\cite{van2020perceptual}.
In \refFig{Overlap} we see this overlap metric (less is better) applied to the result of \methodOur and \methodJitter at different dot counts and for different datasets.
This quantifies what was hinted at before: with \methodJitter, one can get almost-acceptable results as well as very bad ones (as seen by the high variance; now in a while \methodJitter might  discover an accidental \name) while \methodOur is consistently providing a low variance with less overlap.
When dot count increases, variance of \methodJitter becomes less, but the gap to \methodOur becomes even wider.

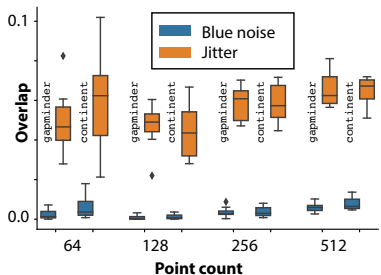
\mywfigure{Overlap}{0.6}{Overlap analysis (see text).}

\mysection{User Study}{UserStudy}
To evaluate \names, we conducted two user studies. The first is a preference study (\refSec{PreferenceStudy}), indicating that \names are considered more appealing over jitter plots. The second is a threshold experiment (\refSec{PerformanceStudy}), confirming that users are performing better to perceive the underlying distribution when using our method.
In both experiments, we compare \methodOur approach (\refAlg{BlueNoisePlots}) to a \methodJitter baseline (\refAlg{JitterPlots}).

\mysubsection{Preference Study}{PreferenceStudy}

\mypara{Methods.}
\change{To evaluate the visual preference, we conducted a user study with a total of $N=12$ participants (3 female, 9 male, $M_{age} = 27.92, SD = 3.26$).
These participants were recruited from a university setting, but without a particular expertise in visualization.}
They were presented with nine different data sets (\dataset{tips, titanic, iris, penguin, geyser, car, gapminder, tooth}, and \dataset{diamond}) visualized using both, \methodOur, as well as \methodJitter treatment, presented in a randomized side-by-side layout.
They were asked two questions: First, to rate which one is ``more visually appealing'' on a choice-enforcing four-point Likert scale, ranging from ``Strongly agree'' to ``Strongly disagree''. Second, to indicate which treatment, if any, they prefer.

\mypara{Analysis.}
Analyzing responses to the first question using a Mann–Whitney U test, we find \methodOur ($Mdn = 2.0$,  $IQR = 1.0$) to be significantly more visually appealing  compared to \methodJitter plots ($Mdn = 1.5$, $IQR = 2.0$, $U = 4091.5$, significant $p < .01$).
Looking at the individual responses, we find clear preferences (significant $p < .05$) for 
\dataset{tips} (difference of \methodJitter and \methodOur of 1.25)
\dataset{penguin} (0.75)
\dataset{iris} (0.50)
\dataset{tooth}	(0.58)
\dataset{gapminder} (0.58)
and lower responses (no significance) for 
\dataset{car} (0.08)
\dataset{titanic} (0.16)	
\dataset{geyser} (0.25) and	
\dataset{diamonds} (0.25).

For the second question we found a preference of \methodOur technique in $62.04\%$ of all responses, in $9.26\%$ of the cases a preference towards the \methodJitter plot and $28.7\%$ without a preference.
When further analysing responses to the second question, for the individual data sets, we find preferences of (\dataset{diamond}: $83.33\%$, \dataset{gapminder}: $75.0\%$, \dataset{geyser}: $75.0\%$, \dataset{penguin}: $83.33\%$, \dataset{tips}: $83.33\%$). These data sets show evenly-distributed points (examples are seen in the~\refFig{JitterVsBN},~\refFig{numPoints}, and~\refFig{Multiclass}, indicating that our approach does support these situations the most. For sparse data sets, participants responded that they do not prefer one of the techniques (\dataset{car}: $58\%$, \dataset{tooth}: $58\%$), possibly due to the fact, that these data sets do not suffer from overdraw.

\mypara{Free-text responses.}
Afterwards, we gave participants the option to respond to the following question using a free text field:``Do you prefer one of the options? If yes, why?''.

While analyzing the free text responses, we found that our participants appreciated the \name not only being ``prettier'', but also for being easier to understand.
They for example stated, that the \name is ``definitely prettier'', it looks ``cleaner and less noisy'', and ``more organized''. Besides this aesthetic aspects, they also stated that the a \name is ``easier to understand'', that dots are ``more easily distinguisable'', and ``easier to count''.

This indicates that our approach might not only be more visually pleasing but also improves the understanding of the data, informing the next study to confirm these subjective judgements.
\change{Further studies would be required to understand preference for variants of our approach, such as centrality or multi-class patterns.}

\mysubsection{Performance Study}{PerformanceStudy}

\mypara{Methods.}
A total of $N=232$ participants from the Amazon MTurk Masters population were simultaneously shown a dot plot on the left and two variants of a distribution to the right (\refFig{Stimulus}).
They were tasked to indicate in a two-alternative forced choice, which variant of the distribution corresponds to the dot plot.
Dot plots were, randomly, either using \methodOur or \methodJitter.
Distributions comprised of B-spline curves with five uniformly-placed control points drawn as line plots.
Their variants result from choosing a random control point in every trial and perturbing it vertically by an offset $O$. 
In every trial, a staircase procedure (QUEST, \cite{watson1983quest}) was conducted to estimate the threshold of $O$ \ie at which level of difference of the reference distribution, different dot plots allow humans to answer correctly in 75\,\% of the cases.
A successful treatment, would have a lower such threshold, which is the dependent variable we record in units of just-noticeable differences (JND)~\cite{ondov2018face,harrison2014ranking,kay2015beyond}.

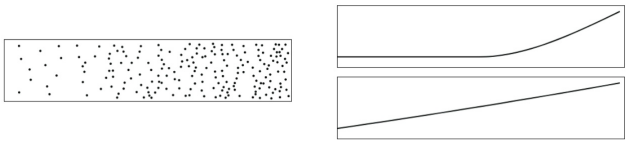
\myfigure{Stimulus}{Experimental stimulus, showing a dot plot of a given data set on the left, and two reference distributions to the right.}

\mypara{Data preparation.}
For 72 participants the threshold experiment did not converge after 100 trials.
In a staircase procedure without bounds this indicates they clicked randomly as any deterministic response will ultimately converge to a value, be it high or low.
Filtering resulted in 160 valid responses.
Based on timings from piloting, participants were paid \$2, for a target rate of \$8/hour.

\mypara{Analysis.}
A Mann-Whitney U test finds a significantly smaller threshold for \names ($Mdn = .34$, $IQR = .16$), compared to the jitter plot ($Mdn = .38$, $IQR = .22$, $U = 2698.0$, $ p = .044$), rejecting the hypothesis that they have identical perceptual thresholds to convey a distribution as a dot plot.

\mypara{Discussion.}
At first, this study design can appear contrived, and it can be asked why not perform a direct comparison. However, there is no single reliable offset $O$ that is valid over all subjects, their viewing conditions, stimuli, training effects, etc.
Hence no $O$ could also be found in a pilot study or using any other process.
Consequently, the difference to study needs to adapt to the conditions, and this is exactly what a staircase procedure does.

Next, one could wonder why JND is a measure of success.
JND is the smallest change a channel (from algorithm over display to the human visual system) encodes.
An efficient visual channel --such as we want our technique to be-- aims to reproduce as many different values as possible, to maximize the entropy, realizing communication with a high bandwidth. Hence, our smaller JNDs indicate the task was made easier as detailed by van Onzenoodt et al.~\cite{ondov2018face}.

\mysection{Discussion}{Discussion}
\mypara{Lloyd relaxation backbone.}
We use Lloyd relaxation as an admittedly simple means to achieve a blue noise spectrum. Many other refined techniques have been proposed\mbox~ \cite{de2012blue,balzer2009capacity,fattal2011blue,schmaltz2010electrostatic,oztireli2012analysis,zhou2012point,leimkuhler2019deep} to produce better blue noise patters, in particular those, that support non-uniform importance. Strikingly, adding data constraints has not found consideration in the literature we are aware of. Lloyd, however, as an expectation maximization process is well suited to enforce constraints iteratively. We further could also use other discretizations of the domain, but this one is particularly amendable to the non-uniform metric used. We hope adding data constraints for visualization purposes could become a new and important sub-task to consider in computer graphics point design methods to come.

\mypara{Point quality.}
We have shown many examples that clearly outperform jitter plots as a baseline found in countless papers printed every day. We further analyzed the dot quality according to state-of-the-art dot correlation metrics. Still the result quality of even the most na\"ive blue noise method could be considered superior to ours, but this is not a plausible comparison to make.
General but biased graphics techniques can remove or add dots, move them freely, etc.\ making the task much easier than our unbiased setting.
But even if there is a gap, it is not clear if the patterns we produce are actually any close to the best patterns we can hope for even with those additional constraints.
A reader is encouraged to apply the blue noise Turing test: is it obvious how to move the points to make the pattern actually better for a human?
We think, yes, maybe, but in many cases only by diminishing returns compared to what is the improvement over jitter.
Future work might find optimization approaches to get point sets that are unbiased in our sense, yet at even higher spectral quality.

\mypara{Visualization impact.}
Hu~\etal~\cite{hu2019data} and Reinert~\etal~\cite{reinert13interactive} have made links between placement of primitives according to data and distribution quality.
Our work is ignorant of the way data points are ultimately presented, so it would be important to have a loop back and ask what size, color, icons or animation would allow for efficient visualization of a dot set, given the spectrum is now high-quality.
In particular, our approach can cover higher dimensions, leading to further visualization questions.
We think both our and their work will open up new problems and solutions in visualization optimizing for aesthetics and clutter avoidance.

\mysection{Conclusions}{Conclusion}
We improve the visual appeal and functionality of jitter plots, by re-casting their randomization into an optimization procedure to put dots ``nicely'', resulting in improved visual appeal and depiction of univariate data sets.
During our user studies, we found that our \name were not only considered to be visually more appealing compared to frequently used jitter plots, but easier to interpret.
Our user study also supports our hypothesis that our plots enable a more accurate estimation of univariate data sets, compared to jitter plots.

\change{
While we use one encoding data dimension and one additional, non-encoding dimension to target the important case of 2D visualization, other combinations are possible.
For 3D \cite{sicat2018dxr} or tangible \cite{lee2012beyond} visualization, an optimization could be extended to fix two data dimensions and optimize a third one.}
In other future work, instances of randomization in visualization, \eg in user interfaces, and Human-Computer interaction, even including the physical world, could be moved forward into  problems where information is neither placed regular, nor random, but inspired by blue noise. 

\bibliographystyle{eg-alpha} 
\bibliography{paper}

\end{document}